\documentclass[twocolumn,prb,aps,floats,showpacs]{revtex4-1}
\usepackage{amsmath}

\usepackage{epsfig}
\usepackage{graphicx}
\usepackage{dcolumn}
\usepackage{bm}
\usepackage{amssymb}
\usepackage[utf8]{inputenc}
\usepackage{color}

\def\gapp{\lower.35em\hbox{$\stackrel{\textstyle>}{\sim}$}}
\def\lapp{\lower.35em\hbox{$\stackrel{\textstyle<}{\sim}$}}

\begin{document}
\bibliographystyle{apsrev4-1}

\title{Power law Kohn anomaly in undoped graphene induced by Coulomb interactions}
\author{F. de Juan}
\author{H.A. Fertig}
\affiliation{Department of Physics, Indiana University, Bloomington, IN 47405, USA}

\date{\today}
\begin{abstract}
Phonon dispersions generically display non-analytic points, known as Kohn anomalies, due to
electron-phonon interactions. We analyze this phenomenon for a zone boundary phonon in
undoped graphene. When electron-electron interactions with coupling constant $\beta$ are taken into
account, one observes behavior demonstrating that the electrons are in a critical phase: the phonon
dispersion and lifetime develop power law behavior with $\beta$ dependent exponents. The observation
of this signature would allow experimental access to the critical properties of the electron
state, and would provide a measure of its proximity to an excitonic insulating phase.
\end{abstract}

\maketitle

\section{Introduction}
The study of electron-electron interactions in graphene, a monolayer of carbon
\cite{CGP09}, remains one of the most interesting open problems in the field and is currently a very
active area of research \cite{KUP10}. The Coulomb interaction plays a very particular
role in this system because its low-energy electronic excitations are described by massless Dirac
fermions. For undoped graphene, the vanishing of the density of states at the Fermi level implies
that the interaction remains truly long-ranged, decaying in real space as $1/r$, and this is
predicted to lead to a number of exotic interaction-induced phenomena, such as logarithmic
renormalization of many physical observables at weak coupling \cite{GGV94,SS07,KUP10}, and
instabilities towards different
symmetry breaking states, like the excitonic insulator \cite{K01,GGG10}, at strong coupling. The
strength of the Coulomb interaction in graphene can be characterized by its bare fine structure
constant $\beta = e^2/\epsilon v_F$, where $v_F$ is the Fermi velocity and $\epsilon$ the dielectric
constant. A naive estimate yields $\beta \approx 2$ for suspended graphene, while lower values are
obtained in the presence of a substrate. This suggests that Coulomb interactions can be relatively
important, but experimentally their strength is still debated \cite{RUY10,EGM11}. 

One of the most striking predicted effects of the Coulomb interaction in graphene is that, because
the Hamiltonian that describes it is scale invariant, some of its correlation functions behave like
power laws with interaction dependendent exponents, so that the system is effectively in a critical
phase \cite{WFM10}. This has also been shown recently by renormalization group arguments
\cite{G10,GMP10}. Similar power laws are also found in Dirac fermion models with
interactions mediated by effective gauge fields \cite{FPS03,GKR03}. Unfortunately most of
the usual experimental probes do not couple to the correlation functions that display
critical behavior, making their experimental observation challenging. 

In this work we show that a signature of this criticality may be accessed
experimentally through the dispersion relation of a zone boundary phonon, the $A_1$ phonon at the
$K$ point. The dispersion of this phonon is produced mainly by its interaction with the Dirac
electrons. Without
electron-electron interactions it shows a square root cusp at $q= q_K \equiv \omega_K/v_F$
($\omega_K$ the phonon frequency at the K point) that crosses over to linear dispersion for
$q>>q_K$, a feature known as a Kohn anomaly. In this work we demonstrate that when the Coulomb
interaction is included, the phonon dispersion is modified strongly: around $q_K$ it becomes a power
law cusp with exponent $\eta(\beta)$, and for $q>>q_K$ it crosses over to another power law
with exponent $\eta_0(\beta)$. The observation of this strong modification of the Kohn
anomaly, in principle feasible with current experimental techniques \cite{MSM07,GSB09,PMF11}, would
provide dramatic evidence of the critical Coulomb interactions in this system, and could potentially
be used as a much needed measurement of their strength $\beta$. This remarkable power law Kohn
anomaly is similar to the one found in some one dimensional systems \cite{Luther}. 

The presence of these powers laws can be understood in simple terms, while their detailed
behavior requires an elaborate calculation discussed below. 
Consider the usual low-energy Hamiltonian for graphene around the $K$ and $K'$ points
\begin{equation}\label{H_el}
H = i v_F \int d^2r \psi^{\dagger} ( \alpha_x \partial_x + \alpha_y \partial_y) \psi,
\end{equation}
with $\vec \alpha = (\tau_z \sigma_x, \sigma_y)$, where the $\sigma$ and $\tau$ matrices act 
on the sublattice and valley degrees of freedom, respectively (spin will be accounted for when
necessary). The chemical potential is set to zero. This Hamiltonian has an SU(2) valley
symmetry generated by the matrices $T_n=(\tau_x
\sigma_y, \tau_y \sigma_y, \tau_z)$, in the sense that the SU(2) rotation $\psi
\rightarrow e^{i T_n
\theta_n}\psi$ leaves the Hamiltonian invariant. 
When the Coulomb interaction
\begin{equation}
H_{int} = \frac{e^2}{2} \int d^2r d^2 r'
\frac{\psi^{\dagger}_r\psi_{r}\psi^{\dagger}_{r'}\psi_{r'}}{|r-r'|}
\end{equation}
is included and for $\beta$ greater than some critical value $\beta_c$, this
system has an instability to an ordered state known as the excitonic insulator \cite{K01,GGG10},
where charge imbalance between sublattices, i.e. an expectation value of $\psi^{\dagger}\sigma_z
\psi$, develops. This instability is reflected in the corresponding susceptibility
$\left<\psi^{\dagger} \sigma_z \psi \psi^{\dagger} \sigma_z \psi\right>$ at $\omega=0$, which
develops a power law $q^{\eta_0}$ with an interaction dependent exponent that goes to zero
for $\beta \rightarrow \beta_c$, signaling the onset of the excitonic phase \cite{WFM10,G10}. 
Therefore, power law behavior in this correlator can be thought of as the weak coupling counterpart
of the excitonic instability, and experimental access to the exponent would allow one to probe how
close
the system is to it. However, this particular susceptibility is difficult to measure, as it requires
a probe that couples differently to the two sublattices. 

The charge density wave instability is however not the only one that Coulomb
interactions can induce. The Hamiltonian (\ref{H_el}) admits two other time-reversal invariant
masses $\tau_x \sigma_x$ and $\tau_y\sigma_x$, and an instability that develops an expectation
value for either of them may proceed in the same way. These order parameters
correspond to a bond density wave order known as the Kekulé distortion \cite{C00}. It can be
shown that the Kekulé and CDW masses, $M_n = (\tau_x \sigma_x, \tau_y \sigma_x, \sigma_z)$ transform
like a spin 1/2 under the valley symmetry, and since the Coulomb
interaction does not break this symmetry, the three instabilities are in fact equivalent: they have
the same weak coupling power law susceptibility with the same exponents.
Since, as we will see, the $A_1$ electron-phonon vertex corresponds to the Kekulé
mass, the phonon self-energy is proportional to the Kekulé susceptibility. Therefore we expect
power law behavior in the phonon dispersion and lifetime. To see this, however, the computation of
the full $\omega$ dependent susceptibility is required. In the remainder of this paper we discuss
the electron-phonon coupling in graphene and the computation of the phonon self-energies with the
aim of establishing precisely where the signatures of critical behavior are to be found. 

\section{Phonons and Kohn anomalies} The phonon spectrum of the honeycomb lattice consists
of six phonon branches, four in-plane and two out-of-plane. Each of these phonons may couple to
electrons near either Dirac point if it has momentum close to zero (a $\Gamma$ point or zone
center phonon), which scatters electrons within each valley, or if it has momentum close to $K$ or
$K'$ points (a zone boundary phonon), in which case it produces intervalley scattering. The
strength of the electron-phonon coupling (EPC), however, depends on how the particular 
displacement pattern of that phonon modifies the hopping integrals between atoms. Two modes
have displacements that produce a significant EPC. The first of these is the
phonon branch of highest energy at the $\Gamma$ point, the $E_2$ phonon.
The second is the $A_1$
branch at the $K$ and $K'$ points (also the highest branch). This is a lattice distortion with a
supercell of six atoms, whose displacement pattern is obtained by taking linear
combinations of the displacements at $K$ and $K'$, and is shown in the inset of Fig
\ref{phononfig}. These two combinations couple to electrons
exactly in the same way as the two components of the Kekulé distortion
\begin{equation}
H_{e-ph,K} =  F_K \int d^2 r \psi^{\dagger} (M_1 u_{K1} + M_2 u_{K2} )\psi, \label{a1H}
\end{equation}
with $F_K = 3 \partial t /\partial a$. For this reason this phonon is also known as the
Kekulé phonon \cite{SA08}. 
The fact that the $E_2$ and $A_1$ phonons are the most predominant is confirmed by Raman
spectroscopy in pristine graphene, where two main peaks are observed \cite{FMS06}; the G peak
corresponds to $E_2$ phonons, while the 2D peak is a second order process involving two $A_1$
phonons. The Hamiltonian of the $A_1$ phonon may be expressed as 
\begin{align}
H = \sum_i \int \frac{d^2q}{(2\pi)^2} \omega_{K} b^{\dagger}_{i,q} b_{i,q},
\end{align}
with creation and destruction operators defined by
\begin{align}
u_i =  \sqrt{\frac{A_c}{4 \omega_K M}} \int \frac{d^2q}{(2\pi)^2} (b_{i,q}e^{i \vec q \vec r} +
b^{\dagger}_{i,q}e^{-i \vec q \vec r}),
\end{align}
where $i=K1,K2$, $\omega_K \approx 0.17$ eV, $A_c$ is the unit cell area. For the range of momenta
$q< 0.25 \AA^{-1}$ where the Dirac fermion model is applicable \cite{CGP09}, the dispersion of the
phonon can be neglected. Indeed phonon band-structure computations excluding the effect of
electron-phonon coupling show a practically flat dispersion \cite{LAW08,KCC09} in this range. A
dimensionless EPC can be defined as 
\begin{equation}
\lambda_K =F_K^2 A_c/(2 M \omega_K v_F^2),
\end{equation}
which is estimated to be in the range $\lambda_{K} \approx 0.03-0.1$ \cite{BA08,BPF09}. 

Due to electron-phonon interactions, phonon dispersion relations are known to develop
non-analytic points, known as Kohn anomalies, at the largest momenta for which the generation of an
electron-hole excitation is kinematically allowed. This renormalization of the dispersion, as well
as the phonon lifetime, can be obtained from the phonon self-energy $\Sigma$, which enters in the
phonon Green's function as
\begin{equation}\label{phonon}
G_{ph}(\omega,q) = \frac{2\omega_K}{\omega^2 - \omega_K^2 -2\omega_K\Sigma(\omega,q)}.
\end{equation}
As anticipated, this self-energy is related to the mass susceptibility, which is
defined as
\begin{equation}
\Pi_{nm} = \left<\psi^{\dagger} M_n \psi \; \psi^{\dagger} M_m \psi\right>,
\end{equation}
because of the form of the coupling given in eq. (\ref{a1H}). The explicit relation follows from the
previous definitions and reads
\begin{equation}
\Sigma = \frac{\lambda_K}{2} \Pi_{11} = \frac{\lambda_K}{2} \Pi_{22}. 
\end{equation}
In the absence of electron-electron interactions, the self-energy can be
computed analytically, and it is given by \cite{PLM04,BPF09}
\begin{equation}\label{nonint} 
\Sigma(\omega,q) = \frac{\lambda_K}{4} (v_F^2 q^2 - \omega^2)^{1/2}.
\end{equation}
Solving for the pole in Eq.
(\ref{phonon}) for small $\lambda_K$, we see the dispersion relation is corrected to
\begin{equation}
\omega(q) \approx \omega_K + \lambda_K/4 (v_F^2 q^2
-\omega_K^2)^{1/2} \label{disprel}
\end{equation}
which has a square root singularity
at $q_K$ for $q>q_K$. For $q<q_K$ the self-energy is purely imaginary, and a finite lifetime is
obtained. The Kohn anomaly is conventionally associated with a linear cusp in the dispersion, which
is obtained only assymptotically for $q>>q_K$; the full dynamical self-energy should be used in
general. Note that $q_K$ is approximately 2\% of the $\Gamma-K$ distance in the Brillouin
zone. The necessity of employing the dynamical
self-energy has been emphasized before \cite{LM06,CG07,THD08}, in particular in the doped case where
the static approximation produces poor agreement with experiments \cite{PLC07}.

\section{Power law mass susceptibility and phonon dispersion} 

\begin{figure}[h]
\begin{center}
\includegraphics[width=8cm]{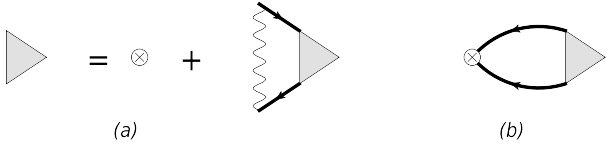}
\caption{(a) Diagrammatic equation for the three point vertex (shaded triangle) in
the ladder approximation. The cross denotes a mass vertex. (b) Response function diagram.
}\label{ladder}
\end{center}
\end{figure}

We will now proceed to compute the general $\omega$ and $q$ dependent mass
susceptibility $\Pi_{nm}$ including the Coulomb interaction. We will see that it acquires
$\beta$ dependent power law behaviour, a feature that is thus inherited by the $A_1$ phonon.
We will employ a ladder summation, as it is the simplest approximation that will capture any
non-analytic behaviour. The ladder summation is represented diagramatically in
Fig. \ref{ladder}. 
Denoting
three-momenta $q=(q_0,\vec q\,)$ one has
\begin{align}\label{response}
\Pi_{nm}(q) = 2i  \int \frac{d^3p}{(2\pi)^3} \; tr \left[M_n G(p) \Gamma_m(p,p+q) G(p+q)\right],
\end{align}
where the mass vertex $\Gamma_m$ is a 4x4 matrix (the sublattice/valley index is
omitted for clarity), and the factor of 2 accounts for spin. In the ladder approximation
$\Gamma_m$ satisfies the
self-consistent equation
\begin{equation}\label{selfc}
\Gamma_m(p,q) = M_m + i e^2 \int \frac{d^3 k}{(2\pi)^2} \frac{G(k)
\Gamma_m(k,q) G(k+q)
}{|\vec p-\vec k|},
\end{equation}
where (we set $v_F=1$ henceforth)
\begin{equation}
G(k)=\frac{k_0 + \vec{\alpha}\vec{k}}{k_0^2-\vec k^2 +i\epsilon}.
\end{equation}
To solve this set of equations, it
is convenient to decompose $\Gamma_m$ in a basis of 4x4 matrices with well defined
transformation
properties under the SU(2) valley symmetry. Defining $\tilde{M}=\tau_z \sigma_z$, this
basis may be taken as the four matrices
$\tilde{M},\mathcal{I},\alpha^i$ which are scalars under this symmetry, and the matrices
$M_n,T_n,\alpha^i T_n$, which transform like a spin 1/2. With this choice we
express $\Gamma_m$ as
\begin{equation}\label{decomp}
\begin{split}
\Gamma_m &= \tilde{\Gamma}_m \tilde{M} + \tilde{\Gamma}^0_m \mathcal{I} + 
\tilde{\Gamma}^i_m \alpha^i \\
& +\Gamma_{nm} M_n+ \Gamma^{0}_{nm} T_n + \Gamma^{i}_{nm} \alpha^i T_n
\end{split}.
\end{equation}
The equations are further
simplified when $\Gamma^{i}_{nm}$ is expressed in terms of its longitudinal and transverse parts
\begin{align}\label{LT}
\Gamma^L_{nm} = \hat q \cdot \vec{\Gamma}_{nm}, &  & 
\Gamma^T_{nm} = \hat q \times \vec{\Gamma}_{nm},
\end{align}
where $\hat q = \vec q / q$, and a similar relation applies for $\tilde{\Gamma}^i_{nm}$. With the
identities
\begin{align}
\vec k \cdot \vec \Gamma_{nm} &= \vec k \cdot \hat q \; \Gamma^L_{nm} - \vec k \times \hat q \;
\Gamma^T_{nm}, \\
\vec k \times \vec \Gamma_{nm} &=\vec k \cdot \hat q \; \Gamma^T_{nm} + \vec k \times \hat q \;
\Gamma^L_{nm},
\end{align}
substituting Eq. (\ref{decomp}) into Eq. (\ref{response}), and
performing the trace, we obtain
\begin{align}\label{response2}
\Pi_{nm}(q) = i  \int & \frac{d^3p}{(2\pi)^3} \;  \frac{8}{D}\left[ f_{11}\Gamma_{nm} + f_{12}
\Gamma^T_{nm}  \right. \nonumber \\ 
& \left. +  \vec p \times \vec q (f_{13}\Gamma^L_{nm} + f_{14}\Gamma^0_{nm}) \right], 
\end{align}
 where we have defined the denominator 
\begin{equation}
D(p,q) = [p_0^2-\vec{p}^2+i\epsilon][(p_0+q_0)^2-(\vec p + \vec
q)^2+i\epsilon],
\end{equation} and where all $f_{ij}(\vec p , \vec q)$ (specified below) are even functions under
the reversal of the
relative angle
$\theta_{\vec p, \vec q} = \theta_p - \theta_q$. Because of the decomposition in Eq. 
(\ref{decomp}), the scalar parts decouple completely and are not needed. We can then obtain
equations for the remaining components of $\Gamma_m$ by multiplying Eq. (\ref{selfc}) by the
corresponding basis matrices and taking the trace. One then obtains
\begin{align}\label{gammanm}
\Gamma_{nm} & = \delta_{nm} -  i\beta \int \frac{d^3 k}{(2\pi)^2}  \frac{1}{D}\frac{1}{|\vec p-\vec
k|}\left[ f_{11}\Gamma_{nm} + f_{12} \Gamma^T_{nm}  \right. \nonumber \\ 
& \left. +  \vec k \times \vec q (f_{13}\Gamma^L_{nm}- f_{14}\Gamma^0_{nm}) \right], \\
\label{gammat}
\Gamma^T_{nm} & = -i\beta \int  \frac{d^3 k}{(2\pi)^2} \frac{1}{D}\frac{1}{|\vec p-\vec
k|}\left[f_{21}\Gamma_{nm}+ f_{22} \Gamma^T_{nm} \right. \nonumber \\ 
& \left. + \vec k \times \vec q \left(f_{23} \Gamma^L_{nm} + f_{24} \Gamma^0_{nm})\right)
\right].
\end{align} 
$\Gamma^L_{mn}$ and $\Gamma^0_{mn}$ satisfy similar
equations, but are not needed in what follows. We now perform a
circular harmonic expansion
\begin{equation}
\Gamma^{(n_p,n_q)} = \int \frac{d\theta_p}{2\pi}e^{i n_p\theta_p}\frac{d\theta_q}{2\pi}e^{i
n_q\theta_q}\Gamma(p, q), 
\end{equation}
and retain only the first order contribution. Terms containing $\vec k \times \vec q$ are odd and
vanish. Thus, $\Gamma^L_n$ and $\Gamma^0_n$ completely decouple to first order.
Moreover, from the structure of Eqs. (\ref{response2}), (\ref{gammanm}) and (\ref{gammat}) it can be
seen that in fact $\Pi_{nm} = \delta_{nm} \Pi$. As expected, the
Kekulé ($\Pi_{11}$ and $\Pi_{22}$) and CDW ($\Pi_{33}$) response functions are the same.

With this simplification the relevant components of $f_{ij}$ are
\begin{align}
f_{11} &=-k_0(k_0+q_0)+ \vec k (\vec k + \vec q\,), \\
f_{12} &=f_{21} =i(\frac{q_0 \vec k \vec q}{q}-k_0q), \\
f_{22} &= \frac{2(\vec q \times \vec k)^2}{q^2}+k_0(k_0+q_0)- \vec k (\vec k + \vec q\,).
\end{align}
Defining the dimensionless kernels
\begin{equation}
K_{ij}^{(n)} = \frac{i}{\pi} \int \frac{d\theta_p}{2\pi} e^{\theta_p n}\int dk_0 k \frac{f_{ij}}{D},
\end{equation}
and 
\begin{equation}\label{coulombker}
C^{(n)}(x) = \int \frac{d \theta_k}{2\pi} \frac{e^{i n \theta_k}}{(1+x^2+ 2x \cos
\theta_k)^{1/2}},
\end{equation}
the self-consistent equations to first order in the circular harmonic expansion finally read
\begin{align}
\Gamma^{(0,0)} = 1+\frac{\beta}{2p}\int dk C^{(0)} (K_{11}^{(0)} \Gamma^{(0,0)} +
K_{12}^{(0)}\Gamma_T^{(0,0)} ), \\
\Gamma_T^{(0,0)} = -\frac{\beta}{2p} \int dk C^{(0)} (K_{21}^{(0)}\Gamma^{(0,0)}+
K_{22}^{(0)}\Gamma_T^{(0,0)}
).\end{align}
A numerical analysis shows that the mixing kernel $K_{12}$ is small compared to $K_{11}$ and may be
neglected also. In this case, the final equations determining the
response function, spelling all momentum dependence, read
\begin{align} 
\Gamma^{(0,0)}(p,q) &= 1+\frac{\beta}{2p}\int_0^{\Lambda} dk C^{(0)}(k/p)
K_{11}^{(0)}(k,q)
\Gamma^{(0,0)}(k,q), \label{finalg} \\
\Pi(q) &= \frac{2}{\pi} \int_0^{\Lambda} d p K_{11}^{(0)}(p,q) \Gamma^{(0,0)}(p,q), \label{finalm}
\end{align}
where $\Lambda$ is an ultraviolet cutoff regularizing the integrals. Note the product $C^{(0)}(k/p)
K_{11}^{(0)}(k,q)$ goes as $p/k$ for large $k$, so the iteration of this equation produces a series
of logarithms characteristic of power law behaviour. Also note that when the external $q<q_0$, all
$K_{ij}$ develop an imaginary part for $(q_0-q)/2 < k < (q_0+q)/2$.

\begin{figure}[h]
\begin{center}
\includegraphics[width=8cm]{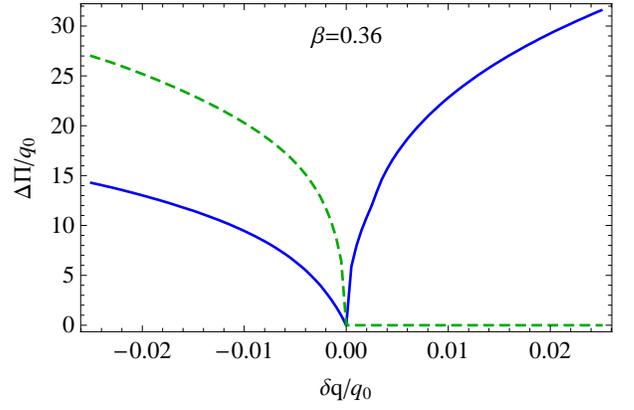}
\caption{Mass susceptibility $\Delta \Pi(q_0,q_0 + \delta q)$ for $|\delta q| <<
q_K$ and $\beta = 0.36$, real part (full line) and imaginary part
(dashed line).}\label{powerlaws}
\end{center}
\end{figure}
\begin{figure}[h]
\begin{center}
\includegraphics[width=8cm]{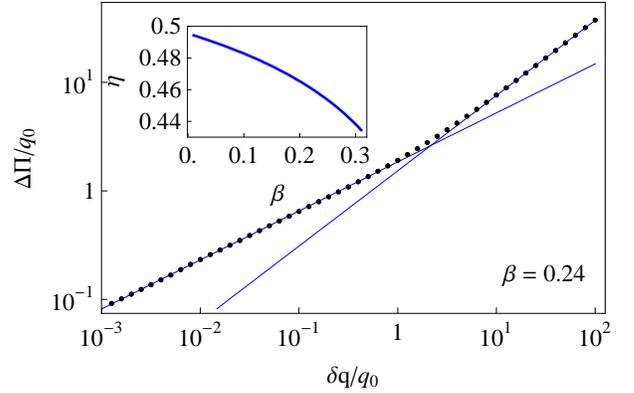}
\caption{Logarithmic plot of the mass susceptibility $\Delta \Pi(q_0,q_0 + \delta q)$ for $\beta =
0.24$ and $\delta q > 0$ (dotted line). The full lines are linear fits
with $\eta=0.45$ for $\delta q << q_K $ and $\eta_0=0.69$ for $\delta q >> q_K$. Inset: The exponent
$\eta$ as a function of $\beta$.}\label{powerlaws2}
\end{center}
\end{figure}

We solve Eq. (\ref{finalg}) numerically by discretizing the momentum $k$
on a logarithmic mesh and solving the corresponding matrix equation by Gaussian elimination. The
integration of Eq. (\ref{finalm}) is straightforward. The result of this procedure is
$\Pi(q_0,q)$. It is convenient to represent it as the difference $\Delta \Pi = \Pi(q_0,q_0 +
\delta q)-\Pi(q_0,q_0)$ with $\delta q = q_0 - q$. Fig. \ref{powerlaws} displays the real and
imaginary parts of $\Delta \Pi$ for $|\delta q| <<
q_0$. We observe a cusp at $\delta q=0$ in the real part, and a finite imaginary part for $\delta
q<0$. Log plots of both sides of the real part and and the imaginary part reveal power
laws as $\delta q \rightarrow 0$. A Kramers-Kronig analysis for $|\delta q| <<
q_0$ shows that this is only consistent if $\Delta \Pi \propto (\delta q)^{\eta}$, i.e. the
exponents are all the same. Fig.
\ref{powerlaws2} shows a log plot for $\delta q>0$
where power law behavior is evident for $\delta q << q_0$. We also observe that $\Delta \Pi$
crosses over to a different power law for $\delta q >> q_0$, which we identify as the static result
$q^{\eta_0}$ \cite{WFM10}. The inset of Fig. \ref{powerlaws2} shows that $\eta$ is
$\beta$-dependent, and that it tends to the non-interacting result in Eq. (\ref{nonint}) as $\beta
\rightarrow 0$. The dependence of $\eta_0$ on $\beta$ can be found in Ref.
\onlinecite{WFM10}.

\begin{figure}[h]
\begin{center}
\includegraphics[width=8.2cm]{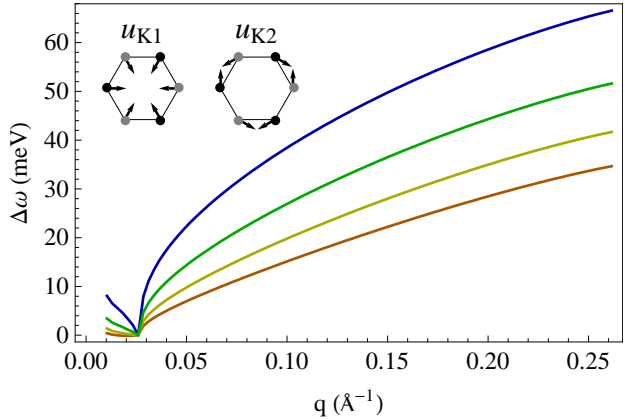}
\caption{$A_1$ phonon dispersion relation $\Delta \omega(q)$ measured from the K point for $\beta =
0,0.1,0.2,0.3$, with higher curves corresponding to higher values of $\beta$. Note that
$\omega(q_K)$, which depends on $\beta$, has been substracted from each curve for an easier
comparison. Inset: the Kekulé phonon displacements.}\label{phononfig}
\end{center}
\end{figure}

Finally, we plot the phonon dispersion relation, which is the main result of this work. This is
given in terms of the self-energy evaluated at the phonon frequency $\omega_K$. To ease the
comparison at different values of $\beta$, we will also represent the difference  
\begin{equation}
\Delta \omega(q) = \omega(q) - \omega(q_K) = \frac{\lambda_K}{2} \left( \Pi(\omega_K,q) -
\Pi(\omega_K,q_K) \right)
\end{equation}
where we have recovered physical units with $\hbar v_F = 6.5$ eV$\AA$. The values of the parameters
used are $\lambda_K=0.1$ and $\Lambda = 1.7 eV$. The phonon dispersion is depicted in Fig.
\ref{phononfig} for different values of $\beta$.  The dispersion follows the
static power law $q^{\eta_0(\beta)}$ for $q>>q_K$, and the cusp turns into $q^{\eta(\beta)}$ as
discussed. 

\section{Discussion} Our computation has shown that interactions turn the Kohn anomaly at the K
point into a power law, so it is natural to ask whether the same effect happens for the anomaly
at $\Gamma$. This is not expected in general grounds, because the
corresponding self-energy is built with vertices corresponding to a conserved
current, and these type of operators do not have anomalous dimensions because of
Ward identities \cite{FPS03}. This is also consistent with the fact that the Coulomb
interaction renormalizes the $A_1$ electron-phonon coupling strongly, but not the $E_2$ one 
\cite{BA08,LAW08}. Power law behavior is thus only expected in the $K$ point anomaly. 

From the experimental point of view, there are several techniques available for the
measurement of
the $A_1$ phonon dispersion, and each one has its own potential difficulties. In general, the power
law at $q>q_K$ appears in a range of momenta that has been already probed with different
techniques, while the cusp structure lies within the precision limits of current experiments, and
may require more effort.

Electron Energy Loss Spectroscopy (EELS) is for example a suitable technique that has already been
used to map the phonon dispersion at the K-point in graphene. This experiments have been
performed on different substrates for which graphene behaves as
quasi-freestanding\cite{YTI05,PMF11}, such as Pt (this is important as hybridization with the
substrate strongly changes the electron band structure and the Kohn anomaly \cite{AW10}). Metallic
screening is however a disadvantage as it spoils the critical behaviour of the electrons, and an
insulating substrate would be more suited to observe the effect. 

A more indirect experiment (with insulating substrate) is to track
the dependence of the 2D Raman peak with incoming laser energy. This method has been used
\cite{MSM07} to measure the dispersion of the $A_1$ phonon. While the amount of data it yields
and the range of momenta it covers is limited and not very close to the $K$ point, the observation
of the $q>q_k$ regime is certainly possible. Finally, X-rays
are a usual tool to measure phonon dispersions in 3D crystals, and while it is probably challenging
to obtain enough intensity from a single sheet of graphene, experiments in graphite
\cite{MET04,GSB09} might be used to deduce the phonon dispersion. This approach is not
straightforward because the electronic structure of graphite is different from graphene, and this
must be taken into account. Nevertheless, it is encouraging to observe that precision measurements
show an $A_1$ phonon dispersion that is not at all linear \cite{GSB09}.

A final comment concerns the robustness of our result to more refined approximations
schemes than the ladder summation. While other sets of diagrams may modify our quantitative
predictions, it is very unlikely that the non-analytic behaviour can be removed in this way. One may
consider, for example, the inclusion of self-energy terms for the electron propagator \cite{GGV94},
which may produce a slow logarithmic dependence of the exponent. Finally, we also note that the 1/N
approximation does gives power law behaviour for the Kekulé mass correlator \cite{G10,GMP10} (and
thus the self-energy) as well.

In summary, this work has shown that the elusive critical behavior of interacting Dirac electrons
in graphene manifests itself through a power law Kohn anomaly for the $A_1$ phonon at the K point.

\section{Acknowledgments} We thank A. Politano for very useful discussions. Support from NSF through
Grant No. DMR-1005035, and from US-Israel Binational Science Foundation (BSF) through Grant No.
2008256 is acknowledged.

\bibliography{kohn}

\begin{thebibliography}{10}%
\makeatletter
\providecommand \@ifxundefined [1]{%
 \ifx #1\undefined \expandafter \@firstoftwo
 \else \expandafter \@secondoftwo
\fi
}%
\providecommand \@ifnum [1]{%
 \ifnum #1\expandafter \@firstoftwo
 \else \expandafter \@secondoftwo
\fi
}%
\providecommand \enquote [1]{``#1''}%
\providecommand \bibnamefont  [1]{#1}%
\providecommand \bibfnamefont [1]{#1}%
\providecommand \citenamefont [1]{#1}%
\providecommand\href[0]{\@sanitize\@href}%
\providecommand\@href[1]{\endgroup\@@startlink{#1}\endgroup\@@href}%
\providecommand\@@href[1]{#1\@@endlink}%
\providecommand \@sanitize [0]{\begingroup\catcode`\&12\catcode`\#12\relax}%
\@ifxundefined \pdfoutput {\@firstoftwo}{%
 \@ifnum{\z@=\pdfoutput}{\@firstoftwo}{\@secondoftwo}%
}{%
 \providecommand\@@startlink[1]{\leavevmode\special{html:<a href="#1">}}%
 \providecommand\@@endlink[0]{\special{html:</a>}}%
}{%
 \providecommand\@@startlink[1]{%
  \leavevmode
  \pdfstartlink
   attr{/Border[0 0 1 ]/H/I/C[0 1 1]}%
   user{/Subtype/Link/A<</Type/Action/S/URI/URI(#1)>>}%
  \relax
 }%
 \providecommand\@@endlink[0]{\pdfendlink}%
}%
\providecommand \url  [0]{\begingroup\@sanitize \@url }%
\providecommand \@url [1]{\endgroup\@href {#1}{\urlprefix}}%
\providecommand \urlprefix [0]{URL }%
\providecommand \Eprint[0]{\href }%
\@ifxundefined \urlstyle {%
  \providecommand \doi [1]{doi:\discretionary{}{}{}#1}%
}{%
  \providecommand \doi [0]{doi:\discretionary{}{}{}\begingroup
  \urlstyle{rm}\Url }%
}%
\providecommand \doibase [0]{http://dx.doi.org/}%
\providecommand \Doi[1]{\href{\doibase#1}}%
\providecommand \bibAnnote [3]{%
  \BibitemShut{#1}%
  \begin{quotation}\noindent
    \textsc{Key:}\ #2\\\textsc{Annotation:}\ #3%
  \end{quotation}%
}%
\providecommand \bibAnnoteFile [2]{%
  \IfFileExists{#2}{\bibAnnote {#1} {#2} {\input{#2}}}{}%
}%
\providecommand \typeout [0]{\immediate \write \m@ne }%
\providecommand \selectlanguage [0]{\@gobble}%
\providecommand \bibinfo [0]{\@secondoftwo}%
\providecommand \bibfield [0]{\@secondoftwo}%
\providecommand \translation [1]{[#1]}%
\providecommand \BibitemOpen[0]{}%
\providecommand \bibitemStop [0]{}%
\providecommand \bibitemNoStop [0]{.\EOS\space}%
\providecommand \EOS [0]{\spacefactor3000\relax}%
\providecommand \BibitemShut [1]{\csname bibitem#1\endcsname}%
\bibitem{CGP09}%
  \BibitemOpen
  \bibfield{author}{%
  \bibinfo {author} {\bibfnamefont{A.~H.}\ \bibnamefont{Castro~Neto}}, \bibinfo
  {author} {\bibfnamefont{F.}~\bibnamefont{Guinea}}, \bibinfo {author}
  {\bibfnamefont{N.~M.~R.}\ \bibnamefont{Peres}}, \bibinfo {author}
  {\bibfnamefont{K.~S.}\ \bibnamefont{Novoselov}},\ and\ \bibinfo {author}
  {\bibfnamefont{A.~K.}\ \bibnamefont{Geim}},\ }%
  \bibfield{journal}{%
  \Doi{10.1103/RevModPhys.81.109}{\bibinfo {journal} {Rev. Mod. Phys.}}\ }%
  \textbf{\bibinfo {volume} {81}},\ \bibinfo {pages} {109} (\bibinfo {year}
  {2009})%
  \bibAnnoteFile{NoStop}{CGP09}%
\bibitem{KUP10}%
  \BibitemOpen
  \bibfield{author}{%
  \bibinfo {author} {\bibfnamefont{V.~N.}\ \bibnamefont{Kotov}}, \bibinfo
  {author} {\bibfnamefont{B.}~\bibnamefont{Uchoa}}, \bibinfo {author}
  {\bibfnamefont{V.~M.}\ \bibnamefont{Pereira}}, \bibinfo {author}
  {\bibfnamefont{A.~H.}\ \bibnamefont{Castro~Neto}},\ and\ \bibinfo {author}
  {\bibfnamefont{F.}~\bibnamefont{Guinea}},\ }%
  \bibinfo {journal} {Rev. Mod. Phys., submitted}%
  \bibAnnoteFile{NoStop}{KUP10}%
\bibitem{GGV94}%
  \BibitemOpen
\bibfield{journal}{%
    }%
  \bibfield{author}{%
  \bibinfo {author} {\bibfnamefont{J.}~\bibnamefont{Gonz\'alez}}, \bibinfo
  {author} {\bibfnamefont{F.}~\bibnamefont{Guinea}},\ and\ \bibinfo {author}
  {\bibfnamefont{M.~A.~H.}\ \bibnamefont{Vozmediano}},\ }%
  \bibfield{journal}{%
  \Doi{10.1016/0550-3213(94)90410-3}{\bibinfo {journal} {Nucl. Phys. B}}\ }%
  \textbf{\bibinfo {volume} {424}},\ \bibinfo {pages} {595} (\bibinfo {year}
  {1994})%
  \bibAnnoteFile{NoStop}{GGV94}%
\bibitem{SS07}%
  \BibitemOpen
  \bibfield{author}{%
  \bibinfo {author} {\bibfnamefont{D.~E.}\ \bibnamefont{Sheehy}}\ and\ \bibinfo
  {author} {\bibfnamefont{J.}~\bibnamefont{Schmalian}},\ }%
  \bibfield{journal}{%
  \Doi{10.1103/PhysRevLett.99.226803}{\bibinfo {journal} {Phys. Rev. Lett.}}\
  }%
  \textbf{\bibinfo {volume} {99}},\ \bibinfo {pages} {226803} (\bibinfo {year}
  {2007})%
  \bibAnnoteFile{NoStop}{SS07}%
\bibitem{K01}%
  \BibitemOpen
  \bibfield{author}{%
  \bibinfo {author} {\bibfnamefont{D.~V.}\ \bibnamefont{Khveshchenko}},\ }%
  \bibfield{journal}{%
  \Doi{10.1103/PhysRevLett.87.246802}{\bibinfo {journal} {Phys. Rev. Lett.}}\
  }%
  \textbf{\bibinfo {volume} {87}},\ \bibinfo {pages} {246802} (\bibinfo {year}
  {2001})%
  \bibAnnoteFile{NoStop}{K01}%
\bibitem{GGG10}%
  \BibitemOpen
  \bibfield{author}{%
  \bibinfo {author} {\bibfnamefont{O.~V.}\ \bibnamefont{Gamayun}}, \bibinfo
  {author} {\bibfnamefont{E.~V.}\ \bibnamefont{Gorbar}},\ and\ \bibinfo
  {author} {\bibfnamefont{V.~P.}\ \bibnamefont{Gusynin}},\ }%
  \bibfield{journal}{%
  \Doi{10.1103/PhysRevB.81.075429}{\bibinfo {journal} {Phys. Rev. B}}\ }%
  \textbf{\bibinfo {volume} {81}},\ \bibinfo {pages} {075429} (\bibinfo {year}
  {2010})%
  \bibAnnoteFile{NoStop}{GGG10}%
\bibitem{RUY10}%
  \BibitemOpen
  \bibfield{author}{%
  \bibinfo {author} {\bibfnamefont{J.~P.}\ \bibnamefont{Reed}}, \bibinfo
  {author} {\bibfnamefont{B.}~\bibnamefont{Uchoa}}, \bibinfo {author}
  {\bibfnamefont{Y.~I.}\ \bibnamefont{Joe}}, \bibinfo {author}
  {\bibfnamefont{Y.}~\bibnamefont{Gan}}, \bibinfo {author}
  {\bibfnamefont{D.}~\bibnamefont{Casa}}, \bibinfo {author}
  {\bibfnamefont{E.}~\bibnamefont{Fradkin}},\ and\ \bibinfo {author}
  {\bibfnamefont{P.}~\bibnamefont{Abbamonte}},\ }%
  \bibfield{journal}{%
  \Doi{10.1126/science.1190920}{\bibinfo {journal} {Science}}\ }%
  \textbf{\bibinfo {volume} {330}},\ \bibinfo {pages} {805} (\bibinfo {year}
  {2010})%
  \bibAnnoteFile{NoStop}{RUY10}%
\bibitem{EGM11}%
  \BibitemOpen
  \bibfield{author}{%
  \bibinfo {author} {\bibfnamefont{D.~C.}\ \bibnamefont{Elias}}, \bibinfo
  {author} {\bibfnamefont{R.~V.}\ \bibnamefont{Gorbachev}}, \bibinfo {author}
  {\bibfnamefont{A.~S.}\ \bibnamefont{Mayorov}}, \bibinfo {author}
  {\bibfnamefont{S.~V.}\ \bibnamefont{Morozov}}, \bibinfo {author}
  {\bibfnamefont{A.~A.}\ \bibnamefont{Zhukov}}, \bibinfo {author}
  {\bibfnamefont{P.}~\bibnamefont{Blake}}, \bibinfo {author}
  {\bibfnamefont{L.~A.}\ \bibnamefont{Ponomarenko}}, \bibinfo {author}
  {\bibfnamefont{I.~V.}\ \bibnamefont{Grigorieva}}, \bibinfo {author}
  {\bibfnamefont{K.~S.}\ \bibnamefont{Novoselov}}, \bibinfo {author}
  {\bibfnamefont{F.}~\bibnamefont{Guinea}},\ and\ \bibinfo {author}
  {\bibfnamefont{A.~K.}\ \bibnamefont{Geim}},\ }%
  \bibfield{journal}{%
  \bibinfo {journal} {Nat. Phys.}}%
   (\bibinfo {year} {2011})%
  \bibAnnoteFile{NoStop}{EGM11}%
\bibitem{WFM10}%
  \BibitemOpen
  \bibfield{author}{%
  \bibinfo {author} {\bibfnamefont{J.}~\bibnamefont{Wang}}, \bibinfo {author}
  {\bibfnamefont{H.~A.}\ \bibnamefont{Fertig}},\ and\ \bibinfo {author}
  {\bibfnamefont{G.}~\bibnamefont{Murthy}},\ }%
  \bibfield{journal}{%
  \Doi{10.1103/PhysRevLett.104.186401}{\bibinfo {journal} {Phys. Rev. Lett.}}\
  }%
  \textbf{\bibinfo {volume} {104}},\ \bibinfo {pages} {186401} (\bibinfo {year}
  {2010})%
  \bibAnnoteFile{NoStop}{WFM10}%
\bibitem{G10}%
  \BibitemOpen
  \bibfield{author}{%
  \bibinfo {author} {\bibfnamefont{J.}~\bibnamefont{Gonz\'alez}},\ }%
  \bibfield{journal}{%
  \Doi{10.1103/PhysRevB.82.155404}{\bibinfo {journal} {Phys. Rev. B}}\ }%
  \textbf{\bibinfo {volume} {82}},\ \bibinfo {pages} {155404} (\bibinfo {year}
  {2010})%
  \bibAnnoteFile{NoStop}{G10}%
\bibitem{GMP10}%
  \BibitemOpen
  \bibfield{author}{%
  \bibinfo {author} {\bibfnamefont{A.}~\bibnamefont{Giuliani}}, \bibinfo
  {author} {\bibfnamefont{V.}~\bibnamefont{Mastropietro}},\ and\ \bibinfo
  {author} {\bibfnamefont{M.}~\bibnamefont{Porta}},\ }%
  \bibfield{journal}{%
  \Doi{10.1103/PhysRevB.82.121418}{\bibinfo {journal} {Phys. Rev. B}}\ }%
  \textbf{\bibinfo {volume} {82}},\ \bibinfo {pages} {121418} (\bibinfo {year}
  {2010})%
  \bibAnnoteFile{NoStop}{GMP10}%
\bibitem{FPS03}%
  \BibitemOpen
  \bibfield{author}{%
  \bibinfo {author} {\bibfnamefont{M.}~\bibnamefont{Franz}}, \bibinfo {author}
  {\bibfnamefont{T.}~\bibnamefont{Pereg-Barnea}}, \bibinfo {author}
  {\bibfnamefont{D.~E.}\ \bibnamefont{Sheehy}},\ and\ \bibinfo {author}
  {\bibfnamefont{Z.}~\bibnamefont{Te\ifmmode \check{s}\else
  \v{s}\fi{}anovi\ifmmode~\acute{c}\else \'{c}\fi{}}},\ }%
  \bibfield{journal}{%
  \Doi{10.1103/PhysRevB.68.024508}{\bibinfo {journal} {Phys. Rev. B}}\ }%
  \textbf{\bibinfo {volume} {68}},\ \bibinfo {pages} {024508} (\bibinfo {year}
  {2003})%
  \bibAnnoteFile{NoStop}{FPS03}%
\bibitem{GKR03}%
  \BibitemOpen
  \bibfield{author}{%
  \bibinfo {author} {\bibfnamefont{V.~P.}\ \bibnamefont{Gusynin}}, \bibinfo
  {author} {\bibfnamefont{D.~V.}\ \bibnamefont{Khveshchenko}},\ and\ \bibinfo
  {author} {\bibfnamefont{M.}~\bibnamefont{Reenders}},\ }%
  \bibfield{journal}{%
  \Doi{10.1103/PhysRevB.67.115201}{\bibinfo {journal} {Phys. Rev. B}}\ }%
  \textbf{\bibinfo {volume} {67}},\ \bibinfo {pages} {115201} (\bibinfo {year}
  {2003})%
  \bibAnnoteFile{NoStop}{GKR03}%
\bibitem{MSM07}%
  \BibitemOpen
  \bibfield{author}{%
  \bibinfo {author} {\bibfnamefont{D.~L.}\ \bibnamefont{Mafra}}, \bibinfo
  {author} {\bibfnamefont{G.}~\bibnamefont{Samsonidze}}, \bibinfo {author}
  {\bibfnamefont{L.~M.}\ \bibnamefont{Malard}}, \bibinfo {author}
  {\bibfnamefont{D.~C.}\ \bibnamefont{Elias}}, \bibinfo {author}
  {\bibfnamefont{J.~C.}\ \bibnamefont{Brant}}, \bibinfo {author}
  {\bibfnamefont{F.}~\bibnamefont{Plentz}}, \bibinfo {author}
  {\bibfnamefont{E.~S.}\ \bibnamefont{Alves}},\ and\ \bibinfo {author}
  {\bibfnamefont{M.~A.}\ \bibnamefont{Pimenta}},\ }%
  \bibfield{journal}{%
  \Doi{10.1103/PhysRevB.76.233407}{\bibinfo {journal} {Phys. Rev. B}}\ }%
  \textbf{\bibinfo {volume} {76}},\ \bibinfo {pages} {233407} (\bibinfo {year}
  {2007})%
  \bibAnnoteFile{NoStop}{MSM07}%
\bibitem{GSB09}%
  \BibitemOpen
  \bibfield{author}{%
  \bibinfo {author} {\bibfnamefont{A.}~\bibnamefont{Gr\"uneis}}, \bibinfo
  {author} {\bibfnamefont{J.}~\bibnamefont{Serrano}}, \bibinfo {author}
  {\bibfnamefont{A.}~\bibnamefont{Bosak}}, \bibinfo {author}
  {\bibfnamefont{M.}~\bibnamefont{Lazzeri}}, \bibinfo {author}
  {\bibfnamefont{S.~L.}\ \bibnamefont{Molodtsov}}, \bibinfo {author}
  {\bibfnamefont{L.}~\bibnamefont{Wirtz}}, \bibinfo {author}
  {\bibfnamefont{C.}~\bibnamefont{Attaccalite}}, \bibinfo {author}
  {\bibfnamefont{M.}~\bibnamefont{Krisch}}, \bibinfo {author}
  {\bibfnamefont{A.}~\bibnamefont{Rubio}}, \bibinfo {author}
  {\bibfnamefont{F.}~\bibnamefont{Mauri}},\ and\ \bibinfo {author}
  {\bibfnamefont{T.}~\bibnamefont{Pichler}},\ }%
  \bibfield{journal}{%
  \Doi{10.1103/PhysRevB.80.085423}{\bibinfo {journal} {Phys. Rev. B}}\ }%
  \textbf{\bibinfo {volume} {80}},\ \bibinfo {pages} {085423} (\bibinfo {year}
  {2009})%
  \bibAnnoteFile{NoStop}{GSB09}%
\bibitem{PMF11}%
  \BibitemOpen
  \bibfield{author}{%
  \bibinfo {author} {\bibfnamefont{A.}~\bibnamefont{Politano}}, \bibinfo
  {author} {\bibfnamefont{A.~R.}\ \bibnamefont{Marino}}, \bibinfo {author}
  {\bibfnamefont{V.}~\bibnamefont{Formoso}},\ and\ \bibinfo {author}
  {\bibfnamefont{G.}~\bibnamefont{Chiarello}},\ }%
  \bibfield{journal}{%
  \Doi{10.1016/j.carbon.2011.09.028}{\bibinfo {journal} {Carbon}}\ }%
  \textbf{\bibinfo {volume} {50}},\ \bibinfo {pages} {734 } (\bibinfo {year}
  {2012})%
  \bibAnnoteFile{NoStop}{PMF11}%
\bibitem{Luther}%
  \BibitemOpen
  \bibfield{author}{%
  \bibinfo {author} {\bibfnamefont{A.}~\bibnamefont{Luther}}\ and\ \bibinfo
  {author} {\bibfnamefont{I.}~\bibnamefont{Peschel}},\ }%
  \bibfield{journal}{%
  \Doi{10.1103/PhysRevB.9.2911}{\bibinfo {journal} {Phys. Rev. B}}\ }%
  \textbf{\bibinfo {volume} {9}},\ \bibinfo {pages} {2911} (\bibinfo {year}
  {1974})%
  \bibAnnoteFile{NoStop}{Luther}%
\bibitem{C00}%
  \BibitemOpen
  \bibfield{author}{%
  \bibinfo {author} {\bibfnamefont{C.}~\bibnamefont{Chamon}},\ }%
  \bibfield{journal}{%
  \Doi{10.1103/PhysRevB.62.2806}{\bibinfo {journal} {Phys. Rev. B}}\ }%
  \textbf{\bibinfo {volume} {62}},\ \bibinfo {pages} {2806} (\bibinfo {year}
  {2000})%
  \bibAnnoteFile{NoStop}{C00}%
\bibitem{SA08}%
  \BibitemOpen
  \bibfield{author}{%
  \bibinfo {author} {\bibfnamefont{H.}~\bibnamefont{Suzuura}}\ and\ \bibinfo
  {author} {\bibfnamefont{T.}~\bibnamefont{Ando}},\ }%
  \bibfield{journal}{%
  \bibinfo {journal} {J. Phys. Soc. Jpn.}\ }%
  \textbf{\bibinfo {volume} {77}},\ \bibinfo {pages} {044703} (\bibinfo {year}
  {2008})%
  \bibAnnoteFile{NoStop}{SA08}%
\bibitem{FMS06}%
  \BibitemOpen
  \bibfield{author}{%
  \bibinfo {author} {\bibfnamefont{A.~C.}\ \bibnamefont{Ferrari}}, \bibinfo
  {author} {\bibfnamefont{J.~C.}\ \bibnamefont{Meyer}}, \bibinfo {author}
  {\bibfnamefont{V.}~\bibnamefont{Scardaci}}, \bibinfo {author}
  {\bibfnamefont{C.}~\bibnamefont{Casiraghi}}, \bibinfo {author}
  {\bibfnamefont{M.}~\bibnamefont{Lazzeri}}, \bibinfo {author}
  {\bibfnamefont{F.}~\bibnamefont{Mauri}}, \bibinfo {author}
  {\bibfnamefont{S.}~\bibnamefont{Piscanec}}, \bibinfo {author}
  {\bibfnamefont{D.}~\bibnamefont{Jiang}}, \bibinfo {author}
  {\bibfnamefont{K.~S.}\ \bibnamefont{Novoselov}}, \bibinfo {author}
  {\bibfnamefont{S.}~\bibnamefont{Roth}},\ and\ \bibinfo {author}
  {\bibfnamefont{A.~K.}\ \bibnamefont{Geim}},\ }%
  \bibfield{journal}{%
  \Doi{10.1103/PhysRevLett.97.187401}{\bibinfo {journal} {Phys. Rev. Lett.}}\
  }%
  \textbf{\bibinfo {volume} {97}},\ \bibinfo {pages} {187401} (\bibinfo {year}
  {2006})%
  \bibAnnoteFile{NoStop}{FMS06}%
\bibitem{LAW08}%
  \BibitemOpen
  \bibfield{author}{%
  \bibinfo {author} {\bibfnamefont{M.}~\bibnamefont{Lazzeri}}, \bibinfo
  {author} {\bibfnamefont{C.}~\bibnamefont{Attaccalite}}, \bibinfo {author}
  {\bibfnamefont{L.}~\bibnamefont{Wirtz}},\ and\ \bibinfo {author}
  {\bibfnamefont{F.}~\bibnamefont{Mauri}},\ }%
  \bibfield{journal}{%
  \Doi{10.1103/PhysRevB.78.081406}{\bibinfo {journal} {Phys. Rev. B}}\ }%
  \textbf{\bibinfo {volume} {78}},\ \bibinfo {pages} {081406} (\bibinfo {year}
  {2008})%
  \bibAnnoteFile{NoStop}{LAW08}%
\bibitem{KCC09}%
  \BibitemOpen
  \bibfield{author}{%
  \bibinfo {author} {\bibfnamefont{S.}~\bibnamefont{Viola~Kusminskiy}},
  \bibinfo {author} {\bibfnamefont{D.~K.}\ \bibnamefont{Campbell}},\ and\
  \bibinfo {author} {\bibfnamefont{A.~H.}\ \bibnamefont{Castro~Neto}},\ }%
  \bibfield{journal}{%
  \Doi{10.1103/PhysRevB.80.035401}{\bibinfo {journal} {Phys. Rev. B}}\ }%
  \textbf{\bibinfo {volume} {80}},\ \bibinfo {pages} {035401} (\bibinfo {year}
  {2009})%
  \bibAnnoteFile{NoStop}{KCC09}%
\bibitem{BA08}%
  \BibitemOpen
  \bibfield{author}{%
  \bibinfo {author} {\bibfnamefont{D.~M.}\ \bibnamefont{Basko}}\ and\ \bibinfo
  {author} {\bibfnamefont{I.~L.}\ \bibnamefont{Aleiner}},\ }%
  \bibfield{journal}{%
  \Doi{10.1103/PhysRevB.77.041409}{\bibinfo {journal} {Phys. Rev. B}}\ }%
  \textbf{\bibinfo {volume} {77}},\ \bibinfo {pages} {041409} (\bibinfo {year}
  {2008})%
  \bibAnnoteFile{NoStop}{BA08}%
\bibitem{BPF09}%
  \BibitemOpen
  \bibfield{author}{%
  \bibinfo {author} {\bibfnamefont{D.~M.}\ \bibnamefont{Basko}}, \bibinfo
  {author} {\bibfnamefont{S.}~\bibnamefont{Piscanec}},\ and\ \bibinfo {author}
  {\bibfnamefont{A.~C.}\ \bibnamefont{Ferrari}},\ }%
  \bibfield{journal}{%
  \Doi{10.1103/PhysRevB.80.165413}{\bibinfo {journal} {Phys. Rev. B}}\ }%
  \textbf{\bibinfo {volume} {80}},\ \bibinfo {pages} {165413} (\bibinfo {year}
  {2009})%
  \bibAnnoteFile{NoStop}{BPF09}%
\bibitem{PLM04}%
  \BibitemOpen
  \bibfield{author}{%
  \bibinfo {author} {\bibfnamefont{S.}~\bibnamefont{Piscanec}}, \bibinfo
  {author} {\bibfnamefont{M.}~\bibnamefont{Lazzeri}}, \bibinfo {author}
  {\bibfnamefont{F.}~\bibnamefont{Mauri}}, \bibinfo {author}
  {\bibfnamefont{A.~C.}\ \bibnamefont{Ferrari}},\ and\ \bibinfo {author}
  {\bibfnamefont{J.}~\bibnamefont{Robertson}},\ }%
  \bibfield{journal}{%
  \Doi{10.1103/PhysRevLett.93.185503}{\bibinfo {journal} {Phys. Rev. Lett.}}\
  }%
  \textbf{\bibinfo {volume} {93}},\ \bibinfo {pages} {185503} (\bibinfo {year}
  {2004})%
  \bibAnnoteFile{NoStop}{PLM04}%
\bibitem{LM06}%
  \BibitemOpen
  \bibfield{author}{%
  \bibinfo {author} {\bibfnamefont{M.}~\bibnamefont{Lazzeri}}\ and\ \bibinfo
  {author} {\bibfnamefont{F.}~\bibnamefont{Mauri}},\ }%
  \bibfield{journal}{%
  \Doi{10.1103/PhysRevLett.97.266407}{\bibinfo {journal} {Phys. Rev. Lett.}}\
  }%
  \textbf{\bibinfo {volume} {97}},\ \bibinfo {pages} {266407} (\bibinfo {year}
  {2006})%
  \bibAnnoteFile{NoStop}{LM06}%
\bibitem{CG07}%
  \BibitemOpen
  \bibfield{author}{%
  \bibinfo {author} {\bibfnamefont{A.~H.}\ \bibnamefont{Castro~Neto}}\ and\
  \bibinfo {author} {\bibfnamefont{F.}~\bibnamefont{Guinea}},\ }%
  \bibfield{journal}{%
  \Doi{10.1103/PhysRevB.75.045404}{\bibinfo {journal} {Phys. Rev. B}}\ }%
  \textbf{\bibinfo {volume} {75}},\ \bibinfo {pages} {045404} (\bibinfo {year}
  {2007})%
  \bibAnnoteFile{NoStop}{CG07}%
\bibitem{THD08}%
  \BibitemOpen
  \bibfield{author}{%
  \bibinfo {author} {\bibfnamefont{W.-K.}\ \bibnamefont{Tse}}, \bibinfo
  {author} {\bibfnamefont{B.~Y.-K.}\ \bibnamefont{Hu}},\ and\ \bibinfo {author}
  {\bibfnamefont{S.}~\bibnamefont{Das~Sarma}},\ }%
  \bibfield{journal}{%
  \Doi{10.1103/PhysRevLett.101.066401}{\bibinfo {journal} {Phys. Rev. Lett.}}\
  }%
  \textbf{\bibinfo {volume} {101}},\ \bibinfo {pages} {066401} (\bibinfo {year}
  {2008})%
  \bibAnnoteFile{NoStop}{THD08}%
\bibitem{PLC07}%
  \BibitemOpen
  \bibfield{author}{%
  \bibinfo {author} {\bibfnamefont{S.}~\bibnamefont{Pisana}}, \bibinfo {author}
  {\bibfnamefont{M.}~\bibnamefont{Lazzeri}}, \bibinfo {author}
  {\bibfnamefont{C.}~\bibnamefont{Casiraghi}}, \bibinfo {author}
  {\bibfnamefont{K.~S.}\ \bibnamefont{Novoselov}}, \bibinfo {author}
  {\bibfnamefont{A.~K.}\ \bibnamefont{Geim}}, \bibinfo {author}
  {\bibfnamefont{A.~C.}\ \bibnamefont{Ferrari}},\ and\ \bibinfo {author}
  {\bibfnamefont{F.}~\bibnamefont{Mauri}},\ }%
  \bibfield{journal}{%
  \Doi{10.1038/nmat1846}{\bibinfo {journal} {Nat. Mater.}}\ }%
  \textbf{\bibinfo {volume} {6}},\ \bibinfo {pages} {198} (\bibinfo {year}
  {2007})%
  \bibAnnoteFile{NoStop}{PLC07}%
\bibitem{YTI05}%
  \BibitemOpen
  \bibfield{author}{%
  \bibinfo {author} {\bibfnamefont{H.}~\bibnamefont{Yanagisawa}}, \bibinfo
  {author} {\bibfnamefont{T.}~\bibnamefont{Tanaka}}, \bibinfo {author}
  {\bibfnamefont{Y.}~\bibnamefont{Ishida}}, \bibinfo {author}
  {\bibfnamefont{M.}~\bibnamefont{Matsue}}, \bibinfo {author}
  {\bibfnamefont{E.}~\bibnamefont{Rokuta}}, \bibinfo {author}
  {\bibfnamefont{S.}~\bibnamefont{Otani}},\ and\ \bibinfo {author}
  {\bibfnamefont{C.}~\bibnamefont{Oshima}},\ }%
  \bibfield{journal}{%
  \Doi{10.1002/sia.1948}{\bibinfo {journal} {Surface and Interface Analysis}}\
  }%
  \textbf{\bibinfo {volume} {37}},\ \bibinfo {pages} {133} (\bibinfo {year}
  {2005})%
  \bibAnnoteFile{NoStop}{YTI05}%
\bibitem{AW10}%
  \BibitemOpen
  \bibfield{author}{%
  \bibinfo {author} {\bibfnamefont{A.}~\bibnamefont{Allard}}\ and\ \bibinfo
  {author} {\bibfnamefont{L.}~\bibnamefont{Wirtz}},\ }%
  \bibfield{journal}{%
  \Doi{10.1021/nl101657v}{\bibinfo {journal} {Nano Letters}}\ }%
  \textbf{\bibinfo {volume} {10}},\ \bibinfo {pages} {4335} (\bibinfo {year}
  {2010})%
  \bibAnnoteFile{NoStop}{AW10}%
\bibitem{MET04}%
  \BibitemOpen
  \bibfield{author}{%
  \bibinfo {author} {\bibfnamefont{J.}~\bibnamefont{Maultzsch}}, \bibinfo
  {author} {\bibfnamefont{S.}~\bibnamefont{Reich}}, \bibinfo {author}
  {\bibfnamefont{C.}~\bibnamefont{Thomsen}}, \bibinfo {author}
  {\bibfnamefont{H.}~\bibnamefont{Requardt}},\ and\ \bibinfo {author}
  {\bibfnamefont{P.}~\bibnamefont{Ordej\'on}},\ }%
  \bibfield{journal}{%
  \Doi{10.1103/PhysRevLett.92.075501}{\bibinfo {journal} {Phys. Rev. Lett.}}\
  }%
  \textbf{\bibinfo {volume} {92}},\ \bibinfo {pages} {075501} (\bibinfo {year}
  {2004})%
  \bibAnnoteFile{NoStop}{MET04}%
\end{thebibliography}%

\end{document}